\documentclass[aps,pra,twocolumn,showpacs]{revtex4}
\usepackage{graphicx,amsmath,amssymb,color}
\usepackage[normalem]{ulem}
\usepackage{amsfonts}
\usepackage[toc,page]{appendix}
\usepackage[ocgcolorlinks,colorlinks=true,linkcolor=blue,citecolor=red]{hyperref}
\usepackage{latexsym}
\usepackage{amsfonts}
\usepackage{algpseudocode}
\usepackage{amsthm}
\usepackage{mathrsfs}
\usepackage{natbib}
\usepackage{color,verbatim}
\usepackage{psfrag}
\DeclareMathAlphabet{\mathrsfs}{U}{rsfs}{m}{n}
\DeclareMathAlphabet{\mathpzc}{OT1}{pzc}{m}{it}
\DeclareMathAlphabet{\matheus}{U}{eus}{m}{n}
\DeclareMathAlphabet{\mathbbold}{U}{bbold}{m}{n}
\DeclareMathOperator{\Tr}{Tr}

\begin{document}

\title{Maximal Non-Classicality in Multi-Setting Bell Inequalities}

\author{Armin Tavakoli$^{1,2}$, Stefan Zohren$^{3}$, Marcin Pawlowski$^{2}$ }
\affiliation{$^1$Department of Physics, Stockholm University,
SE-10691 Stockholm, Sweden\\
$^2$Institute of Theoretical Physics and Astrophysics, Uniwersytet Gda\'nski, PL-80-952 Gda\'nsk, Poland\\
$^3$Quantum and Nanotechnology Theory Group, Department of Materials, and\\
Machine Learning Research Group, Department of Engineering Science, University of Oxford, UK}


\date{\today}


\begin{abstract}
The discrepancy between maximally entangled states and maximally non-classical quantum correlations is well-known but still not well understood. We aim to investigate the relation between quantum correlations and entanglement in a family Bell inequalities with $N$-settings and $d$ outcomes. Using analytical as well as numerical techniques, we derive both maximal quantum violations and violations obtained from maximally entangled states. Furthermore, we study the most non-classical quantum states in terms of their entanglement entropy for large values of $d$ and many measurement settings. 
Interestingly, we find that the entanglement entropy behaves very differently depending on whether $N=2$ or $N> 2$: when $N=2$ the entanglement entropy is a monotone function of $d$ and the most non-classical state is far from maximally entangled, whereas when $N> 2$ the entanglement entropy is a non-monotone function of $d$ and converges to that of the maximally entangled state in the limit of large $d$.
\end{abstract}


\pacs{03.67.Hk,
03.67.-a,
03.67.Dd}

\maketitle

\section{Introduction}
Quantum correlations arising in space-like separated measurement events can be subject to strong correlations that cannot be explained by classical physics \cite{B64}. The strength of classical correlations is bounded by Bell inequalities, which are known to have quantum violations, certifying the non-classicality of the physics at play. A necessary condition to violate a Bell inequality is sharing an entangled quantum state \cite{HHHH09}. For pure states, the standard measure of entanglement is the entanglement entropy \cite{BBP96}, defined as the von Neumann entropy of a subsystem of the state. The relation between entanglement and the strength of quantum correlations has long been considered an interesting question. 

For the simplest Bell scenario, with two parties Alice and Bob performing one of two two-outcome measurements, the CHSH inequality \cite{CHSH69} constitutes a tight bound on the set of classical correlations \cite{F82}. To maximally violate the CHSH inequality, one must distribute a state with maximal entanglement entropy (the so-called maximally entangled state). However, when the CHSH inequality was generalized to the CGLMP inequality \cite{CGLMP02} which considers scenarios with Alice and Bob choosing one of two $d$-outcome measurements, it was discovered that the maximal quantum violation of the CGLMP inequality for some low values of $d>2$ cannot be achieved by distributing a maximally entangled state between Alice and Bob \cite{AGG05, ADGL02}. The CGLMP inequality has been studied for scenarios with number of outcomes ranging up to the order of $10^6$, and the most non-classical state is found to be far from maximally entangled \cite{ZG08}. Recent results \cite{FP15} suggest that the discrepancy is due to the chosen measures quantifying quantum correlations.

Quantum correlations are relevant in Bell scenarios beyond the scope of the CGLMP inequality. Significant research effort has been directed at Bell scenarios where Alice and Bob can choose between more than two measurements. Such scenarios with two-outcome measurements for Alice and Bob have been explored in e.g.\ Refs.\ \cite{joinedreference, CG04} and many-outcome scenarios in e.g.\ Refs.\ \cite{CG04, joinedreference2}. In addition, Bell scenarios with many settings and more than two parties have been studied in \cite{L08}.

Here we present a much simplified version of the multi-setting generalization of the CGLMP inequality derived in Ref.\ \cite{BKP06} i.e.\ for scenarios where Alice and Bob perform one of $N$ $d$-outcome measurements. We investigate quantum violations of this inequality for arbitrary $N$ and $d$, using both maximally entangled states and general entangled states. In addition to the maximal violations of the inequality, we study the entanglement entropy of the most non-classical quantum state by both analytical and numerical means.

Interestingly, we find that the nature of the most non-classical state is very different depending on whether $N=2$ (the CGLMP inequality) or $N>2$. When $N=2$, the entanglement entropy of the most non-classical state is a monotonically decreasing function of $d$ and thus as we increase $d$ the state becomes less entangled. However, we find that when $N>2$ the entanglement entropy is a non-monotone function of $d$ and that the entanglement entropy of the most non-classical state converges to that of the maximally entangled state in the limit of large $d$. 

\section{Multi-setting and many outcome Bell inequalities}
Let Alice and Bob share a bipartite quantum state of local Hilbert space dimension $D$: $|\psi\rangle\in \mathbb{C}^D\otimes \mathbb{C}^D$. Alice and Bob can each perform one of $N$ measurements, indexed by $x,y=0,...,N-1$ respectively, with associated outcomes $A_x,B_y$ each of which can take $d$ different values, where $d\leq D$, on some ordered space. We will focus on cases where the number of outcomes equals the local Hilbert space dimension; that is when $D=d$. The constraint $D=d$ will from now on be assumed unless otherwise is stated. Scenarios with $D>d$ will be briefly discussed. 

For the given Bell scenario, we can construct the following Bell inequality,
\begin{multline}\label{Ineq}
B_{N,d}\equiv P(A_{N-1}\geq B_0)+\sum_{n=0}^{N-1}P(A_n<B_n) +\\
+ \sum_{n=0}^{N-2}P(B_{n+1}<A_n)\geq 1.
\end{multline}

\begin{figure}
\centering
\includegraphics[width=0.6\columnwidth]{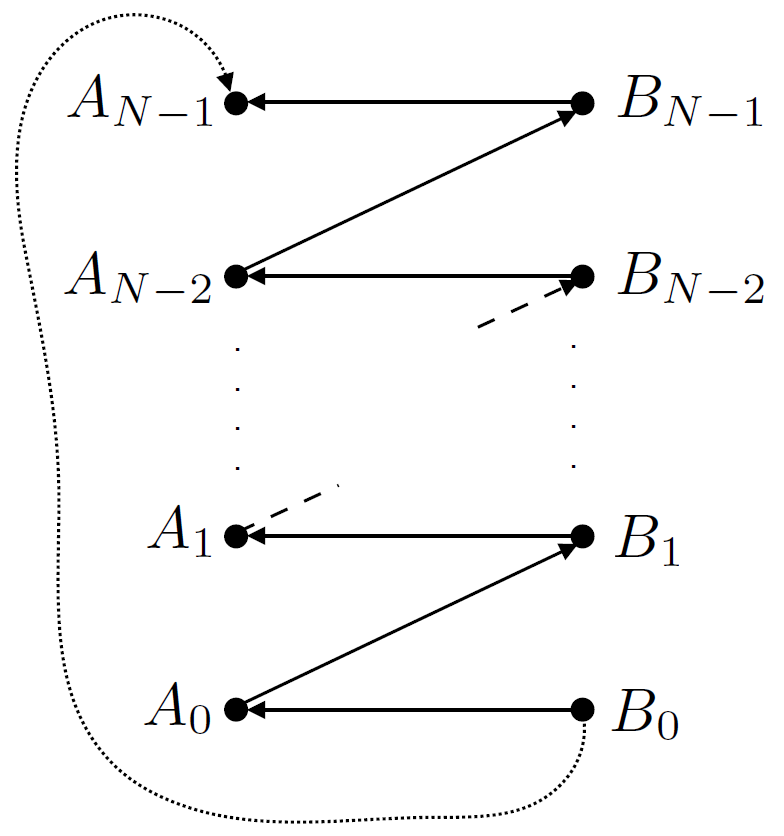}
\caption{A graphical illustration of the chaining associated with the inequality \eqref{Ineq} where the arrows represent the relation `$<$' between two measurement outcomes.}
\label{fig:Ineq}
\end{figure}

The inequality \eqref{Ineq} is a chained Bell inequality, and we illustrate it graphically in Figure \ref{fig:Ineq}.

To prove this assertion, we extend the method of Ref.\ \cite{ZG08}. First note the following fact: $\{A_{N-1}\geq B_{N-1}\}\cap\{B_{N-1}\geq A_{N-2}\}\cap\ldots\cap \{B_1\geq A_0\}\cap \{A_0\geq B_0\} \subset \{A_{N-1}\geq B_0\}$. Therefore, if we take the complement of both sides, it holds that $\{A_{N-1}<B_{0}\}\subset \{A_{N-1}< B_{N-1}\}\cup \{B_{N-1}<A_{N-2}\}\cup\ldots\cup \{B_1< A_0\}\cup \{A_0<B_0\}$ (see Figure \ref{fig:Ineq}). Any classical probability distribution in the given Bell scenario takes the form of a convex mixture of fully deterministic probability distributions, for which all probabilities are trivial i.e.\ either zero or one. Therefore, since randomness is not a fundamental property but a mere statistical property in classical probability distributions, it must hold that $P(A_{N-1}<B_0)=1-P(A_{N-1}\geq B_0)\leq P(A_0< B_0)+P(B_1<A_0)+\ldots+P(B_{N-1}<A_{N-2})+P(A_{N-1}<B_{N-1})$, from which the inequality \eqref{Ineq} follows.

Our inequality is (up to a permutation of parties and measurement settings) a simplified version of the inequality derived in \cite{BKP06}, analogous to the simplification of the CGLMP inequality in Ref.\ \cite{ZG08}. An advantage of our inequality is that its form is independent of $d$. Another advantage is that the inequality only depends on the relative ordering of the outcomes, and not their ascribed values. 

For sake of completeness, we show that of the specific case when the outcomes take the values $0,...,d-1$, our inequality is equivalent to that of Ref. \cite{BKP06} (again, up to a permutation of parties and measurement settings), namely 
\begin{multline}\label{Barret}
I_{N,d}\equiv \langle [B_0-A_0]\rangle+\langle [A_0-B_1]\rangle+...+\langle [B_{N-1}-A_{N-1}]\rangle\\
+\langle [A_{N-1}-B_0-1]\rangle\geq d-1
\end{multline}
where $\langle X\rangle =\sum_{l=0}^{d-1}lP(X=l)$ and $[X]=X \mod{d}$. Simply note that we can write $[X]=X-d\lfloor X/d\rfloor$ which implies that the inequality can be written $\big\langle -1-d\big(\sum_{n=0}^{N-1} \lfloor (B_n-A_n)/d\rfloor +\sum_{n=0}^{N-2}\lfloor (A_n-B_{n+1})/d\rfloor+ \lfloor (A_{N-1}-B_0-1)/d\rfloor\big)\big\rangle\geq d-1$. If we use that $A_i,B_i\in\{0,...,d-1\}$, the inequality \eqref{Ineq} is implied.

\section{Quantum Violations}

We now investigate the quantum properties of the Bell expression $B_{N,d}$.  We consider both violations using maximally entangled states i.e.\ a state of the form $|\psi_{max}\rangle=1/\sqrt{d}\sum_{k=0}^{d-1}|kk\rangle$, and non-maximally entangled states. 

Let us denote the outcomes of Alice's and Bob's measurements by $a,b\in\{0,...,d-1\}$. In Ref.\ \cite{BKP06}, particular measurements were introduced with the motivation that the maximally entangled state has the property that if Alice measures an observable with an eigenstate $|s\rangle$, and Bob measures an observable with conjugated eigenvectors $|s\rangle^*$, correlations will be perfect. Therefore, we let Alice and Bob measure in the respective bases given by
\begin{eqnarray}\label{Meas1}
|a\rangle_{A,x}=\frac{1}{\sqrt{d}}\sum_{k=0}^{d-1}\omega^{k(a+\alpha_x^{(N)})}|k\rangle\\\label{Meas2}
|b\rangle_{B,y}=\frac{1}{\sqrt{d}}\sum_{k=0}^{d-1}\omega^{k(-b+\beta_y^{(N)})}|k\rangle
\end{eqnarray}
where $\alpha_x^{(N)}=x/N$ and $\beta_y^{(N)}=(1-2y)/2N$ for $x,y=0,...,N-1$, and $\omega=e^{i\frac{2\pi}{d}}$. 

We have conducted extensive numerical studies using semidefinite programs (SDPs) \cite{VB96} to investigate whether these measurements are optimal. However, the problem of optimizing the quantity $B_{N,d}$ over the measurement of Alice, the measurements of Bob, and the shared state, cannot be directly maped into an SDP. Instead we have used a see-saw technique in which we first use an SDP to optimize only over the measurements of Alice. Then we run another SDP to optimize over the measurements of Bob while fixing Alice's measurements as obtained in the previous SDP. Finally, we fix Alice's and Bob's measurements as returned by the two previous SDPs and run a third SDP optimizing over only over the shared state. This procedure is iterated until a stable value of $B_{N,d}$ is obtained. Naturally, this procedure does not guarantee that $B_{N,d}$ achieves its global optimum under quantum theory. Yet, for all investigated cases ranging over several low values of $N,d$, the obtained value of $B_{N,d}$ by the see-saw method returns the above measurements of Alice and Bob. Due to this numerical evidence, we will conjecture the above measurements as optimal.  

The numerics supporting the conjecture of the optimal measurements is performed with the constraint that $D=d$. We have used SDPs in an analogous see-saw technique to investigate the value of $B_{N,d}$ in various scenarios when $D>d$. Specifically, we have considered $N=2,3,4,5$, $d=2,...,9$ and $D=d,...,10$. We have observed no instance in which $D>d$ yields an enhancement over the case $D=d$.

Following the conjecture of the optimal measurements, we can define the Bell operator, $\mathscr{B}_{N,d}$, associated with the Bell expression $B_{N,d}$ as
\begin{multline}
\mathscr{B}_{N,d}\equiv \sum_{a\geq b} |a\rangle_{A,N-1}\langle a|_{A,N-1}\otimes |b\rangle_{B,0}\langle b|_{B,0}\\
+\sum_{n=0}^{N-1}\sum_{b>a}|a\rangle_{A,n}\langle a|_{A,n}\otimes |b\rangle_{B,n}\langle b|_{B,n}\\
+\sum_{n=0}^{N-2}\sum_{a>b}|a\rangle_{A,n}\langle a|_{A,n}\otimes |b\rangle_{B,n+1}\langle b|_{B,n+1}.
\end{multline}
Thus, we can write that $B_{N,d}=\Tr\left(\rho \mathscr{B}_{N,d}\right)$ where $\rho$ is some shared quantum state. The optimal value of $B_{N,d}$ is found from computing the smallest eigenvalue of $\mathscr{B}_{N,d}$, and the corresponding eigenvector is the associated quantum state that gives rise to the maximal violation i.e.\ it is the most non-classical quantum state. 

However, in the case of pure quantum states we can extensively simplify the eigenvalue computation by using the Schmidt decomposition to write any pure quantum state in the form $|\phi\rangle=\sum_{k=0}^{d-1}\lambda_k|kk\rangle$ for $\lambda_k$ being real numbers respecting the normalization $\sum_{k=0}^{d-1}\lambda_k^2=1$. It is then possible to write the Bell expression $B_{N,d}$ in the following form 
\begin{equation}\label{Sum}
B_{N,d}=\sum_{k,l=0}^{d-1}M_{kl}\lambda_k\lambda_l
\end{equation}
where the $d\times d$ matrix $M$ is given by
\begin{equation}\label{M}
M_{kl}=N\delta_{kl}-\frac{N}{d}\frac{\sin\left(\frac{(N-1)\pi (k-l)}{dN}\right)}{\sin\left(\frac{\pi (k-l)}{d}\right)}.
\end{equation}
This expression is derived in appendix \ref{AppA}. 

The problem of finding the maximal quantum violation of the inequalities reduces to finding the smallest eigenvalue of $M$. However, we will first investigate optimal values of $B_{N,d}$ achievable by performing measurements on a shared maximally entangled state.

\subsection{Violations from maximally entangled states}

The maximally entangled state, $|\psi_{max}\rangle$ has Schmidt coefficients $\lambda_{k}= 1/\sqrt{d}$ for $k=0,...,d-1$. Inserting this into \eqref{Sum}, the value of the Bell expression is easily computed from
\begin{equation}\label{sument}
B_{N,d}(|\psi_{max}\rangle)=\frac{1}{d}\sum_{k,l=0}^{d-1}M_{kl}.
\end{equation}
Let us study the limiting case of many measurements, i.e.\ find the value of $B_{\infty,d}(|\psi_{max}\rangle)\equiv \lim_{N\rightarrow \infty} B_{N,d}(|\psi_{max}\rangle)$. 
We first compute the probability distribution arising from Alice and Bob making measurements \eqref{Meas1} and \eqref{Meas2} on $|\psi_{max}\rangle$. This can be obtained directly from equation \eqref{Prob1}:
\begin{equation}\label{Prob2}
\!\!\!\!P_{N,d}(a,b|x,y)\!=\!\!\frac{1}{d^3}\frac{\sin^2\left(\pi\left(a-b+\alpha_x^{(N)}+\beta_y^{(N)}\right)\right)}{\sin^2\left(\frac{\pi}{d}\left(a-b+\alpha_x^{(N)}+\beta_y^{(N)}\right)\right)}.
\end{equation}
Using the probability distribution, we can expand the Bell expression $B_{N,d}(|\psi_{max}\rangle)$ in \eqref{Ineq} as
\begin{multline}\label{B}
B_{N,d}(|\psi_{max}\rangle)=\\ \frac{\sin^2\left(\frac{\pi}{2N}\right)}{d^2\sin^2\left(\frac{\pi}{d}\left(1-\frac{1}{2N}\right)\right)}+\frac{1}{d^3}\sum_{j=1}^{d-1}(d-j)\sin^2\left(\frac{\pi}{2N}\right)\\
\times\left[\frac{2N-1}{\sin^2\left(\frac{\pi}{d}\left(j-\frac{1}{2N}\right)\right)}+\frac{1}{\sin^2\left(\frac{\pi}{d}\left(j+1-\frac{1}{2N}\right)\right)}\right].
\end{multline}
In the limit of large $N$, it is evident that there is a contribution to $B_{\infty,d}(|\psi_{max}\rangle)$ only from the final term in the sum associated to $j=d-1$. Using Taylor expansion of the sine function, we find that the limit is 
\begin{equation}\label{lim}
B_{\infty, d}(|\psi_{max}\rangle)=\frac{1}{d},
\end{equation}
which goes to zero as $d\to\infty$. In fact, the right-hand-side of \eqref{lim} is the no-signaling bound of $B_{N,d}$. This is  realized from the fact that the no-signaling bound of $I_{N,d}$ in \eqref{Barret} coincides with its algebraic bound, that is zero, and that the affine transformation relating $B_{N,d}$ and $I_{N,d}$ is $I_{N,d}=d\times B_{N,d}-1$ up to permutations of parties and settings. Thus, in the limit of many measurement settings, the above result of quantum theory does converge to the no-signaling bound. However, the algebraic bound of $B_{N,d}$ is zero, and thus there is a discrepancy between the no-signaling bound and the algebraic bound of $B_{N,d}$.

In the other limiting case of many outcomes, finding the value of  $B_{N,\infty}(|\psi_{max}\rangle)\equiv \lim_{d\rightarrow \infty} B_{N,d}(|\psi_{max}\rangle)$ is a bit more involved. We can use the fact that the sum in equation \eqref{sument} can be evaluated as an integral, which can be related to the Trigamma function,  $\psi_{1}(z)$, defined as the second logarithmic derivative of the Gamma function, $\psi_1(z)=\frac{d^2}{dz^2}\log \Gamma(z)$, as follows
\begin{multline}\label{Int}
B_{N,\infty}(|\psi_{max}\rangle)=
N-N\int_{0}^{1}\int_{0}^{1}\frac{\sin\left(\frac{(N-1)\pi (x-y)}{N}\right)}{\sin\left(\pi (x-y)\right)}dxdy\\
=\frac{2N}{\pi^2}\psi_{1}\left(1-\frac{1}{2N}\right)\sin^2\left(\frac{\pi}{2N}\right).
\end{multline}
In appendix \ref{AppB}, we prove this relation. It is worth noting that for the special case of $N=2$, it was shown in \cite{ZG08} that the limit evaluates to $B_{2,\infty}(|\psi_{max}\rangle)=2-16 \text{Cat}/\pi^2$ where $\text{Cat}$ is Catalan's constant, corresponding to what was obtained for the CGLMP inequality in Ref.\ \cite{CGLMP02}.

Evidently, unless we let both $N,d\rightarrow \infty$, we cannot reach the algebraic bound of $B_{N,d}$ by using maximally entangled states. This puts our inequality in contrast to that of Ref.\cite{BKP06} in which the algebraic bound coincides with the no-signaling bound and thus can be asymptotically achieved in the limit $N\rightarrow \infty$.

\subsection{Optimal quantum violations}
We will now investigate the optimal quantum violations of the inequalities \eqref{Ineq} by sharing general entangled states. As we have already mentioned, the optimal quantum violation of the inequality corresponds to the smallest eigenvalue of the matrix $M$ and the entangled state giving rise to the violation is the associated eigenvector of $M$. This reasoning relies on the measurements \eqref{Meas1} and \eqref{Meas2} indeed being optimal choices. To strengthen this conjecture, we have verified the optimality of these results numerically using SDPs \cite{VB96} for the cases $N,d=2,...,10$.

\begin{figure}
\centering
\includegraphics[width=1\columnwidth]{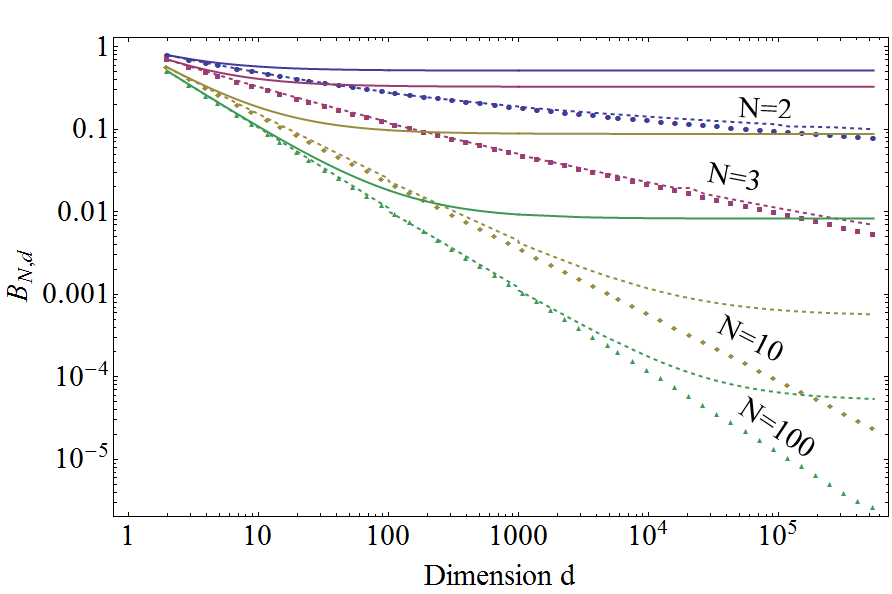}
\caption{The optimal quantum violations for $N=2,3,10,100$ and dimensions up to about $d=0.5\times 10^6$. Solid lines are obtained from maximally entangled states, the dotted lines from the most non-classical quantum state, and the dashed lines from the approximation state $|\Phi_{app}^{(N)}\rangle$ in \eqref{Approx}.}
\label{fig:1}
\end{figure}

In Figure \ref{fig:1}, we have computed the maximal violation of the inequalities, the violation using a maximally entangled state, and also the violation due to the approximate state $|\Phi_{app}^{(N)}\rangle$ which is discussed below. We have chosen some particular values of $N=2,3,10,100$ and $d$ ranging from $d=2$ up to approximately about $d=0.5\times 10^6$ in exponentially increasing increments. The computation of the smallest eigenvalue of $M$ (the maximal violation) can be efficiently implemented using power iteration over the Krylov space associated to $M$. We elaborate on the numerical technique in appendix \ref{AppC}.

The results suggest that the optimal violation of the inequality approaches the algebraic bound of $B_{N,d}$, namely zero, as $d\rightarrow \infty$ for all investigated values of $N$. The convergence to the algebraic bound is more rapid with increasing $N$.

\subsection{Approximating the most non-classical state}
Due to the apparent discrepancy between maximally entangled state and the most non-classical state, $|\psi_{opt}\rangle$, in terms of their ability to minimize $B_{N,d}$, it is interesting to investigate the properties of $|\psi_{opt}\rangle$.

For simplicity, if we represent $|\psi_{opt}\rangle $ in the Schmidt basis as $|\psi_{opt}\rangle=\sum_{k=0}^{d-1}\lambda_k|kk\rangle$, it follows from the symmetries of the matrix $M$ that $|\psi_{opt}\rangle$  is subject to the symmetry property that $\lambda_k=\lambda_{d-1-k}$ whenever $d\geq 2$. However, despite being easy to compute for given values $(N,d)$, the exact form of $|\psi_{opt}\rangle $ is difficult to find by analytical means.

However, we propose an approximation of the state $|\psi_{opt}\rangle$  for arbitrary $N$ and $d$, which denote by $|\Phi_{app}^{(N)}\rangle$:
\begin{equation}\label{Approx}
|\Phi_{app}^{(N)}\rangle=\frac{1}{\sqrt{ \mathscr{N}}} \sum_{k=0}^{d-1}\frac{|kk\rangle}{\left[(k+1)(d-k)\right]^{1/N}}
\end{equation}
where $\mathscr{N}=\sum_{j=0}^{d-1}\left[(j+1)(d-j)\right]^{-\frac{2}{N}}$ is the normalization. 

The special case amounting to $N=2$ was initially proposed in Ref.\ \cite{CC06} to approximate the state maximally violating the CGLMP inequality. The state $|\Phi_{app}^{(2)}\rangle$ was extensively studied for dimensions up to the order of $10^6$ in Ref.\ \cite{ZRG10}. On the one hand, it was shown that $|\Phi_{app}^{(2)}\rangle$ yields a good approximation of the optimal violation of the CGLMP inequality, especially for $d<1000$, and that it yields a violation which converges to the algebraic bound in the limit of many outcomes \cite{ZRG10}. On the other hand, the numerics of Ref.\ \cite{ZRG10} shows that $|\Phi_{app}^{(2)}\rangle$ is quite a bad approximation of the state $|\psi_{opt}\rangle$ in terms of entanglement entropy. Although it was shown that the entropy of $|\Phi_{app}^{(2)}\rangle$ becomes $E(|\Phi_{app}^{(2)}\rangle)=1/2$ in the limit of many outcomes, where it converges to the most non-classical state \cite{ZRG10}.  For general $N$, the numerics in Figure \ref{fig:1} indicate that the violation due to the approximate state $|\Phi_{app}^{(N)}\rangle$ converges to the algebraic bound of $B_{N,d}$ in the limit of many outcomes. Furthermore, it appears that with higher $N$, the violation due to $|\Phi_{app}^{(N)}\rangle$ increasingly well approximates the optimal quantum violation of the inequality.

The question of how well the state $|\Phi_{app}^{(N)}\rangle$ approximates the entanglement entropy of the most non-classical state $|\psi_{opt}\rangle$ is investigated in the next section.

\section{Maximal non-classicality versus entanglement}
In order to investigate the relation between maximal non-classicality and entanglement, we will  study the entanglement entropy of the state $|\psi_{opt}\rangle$ by extensive numerics for high $N$ and $d$. We will also study to what extent the degree of entanglement of $|\psi_{opt}\rangle$ can be approximated with the introduced state $|\Phi^{(N)}_{app}\rangle$. 

\begin{figure}[t]
\centering
\includegraphics[width=1\columnwidth]{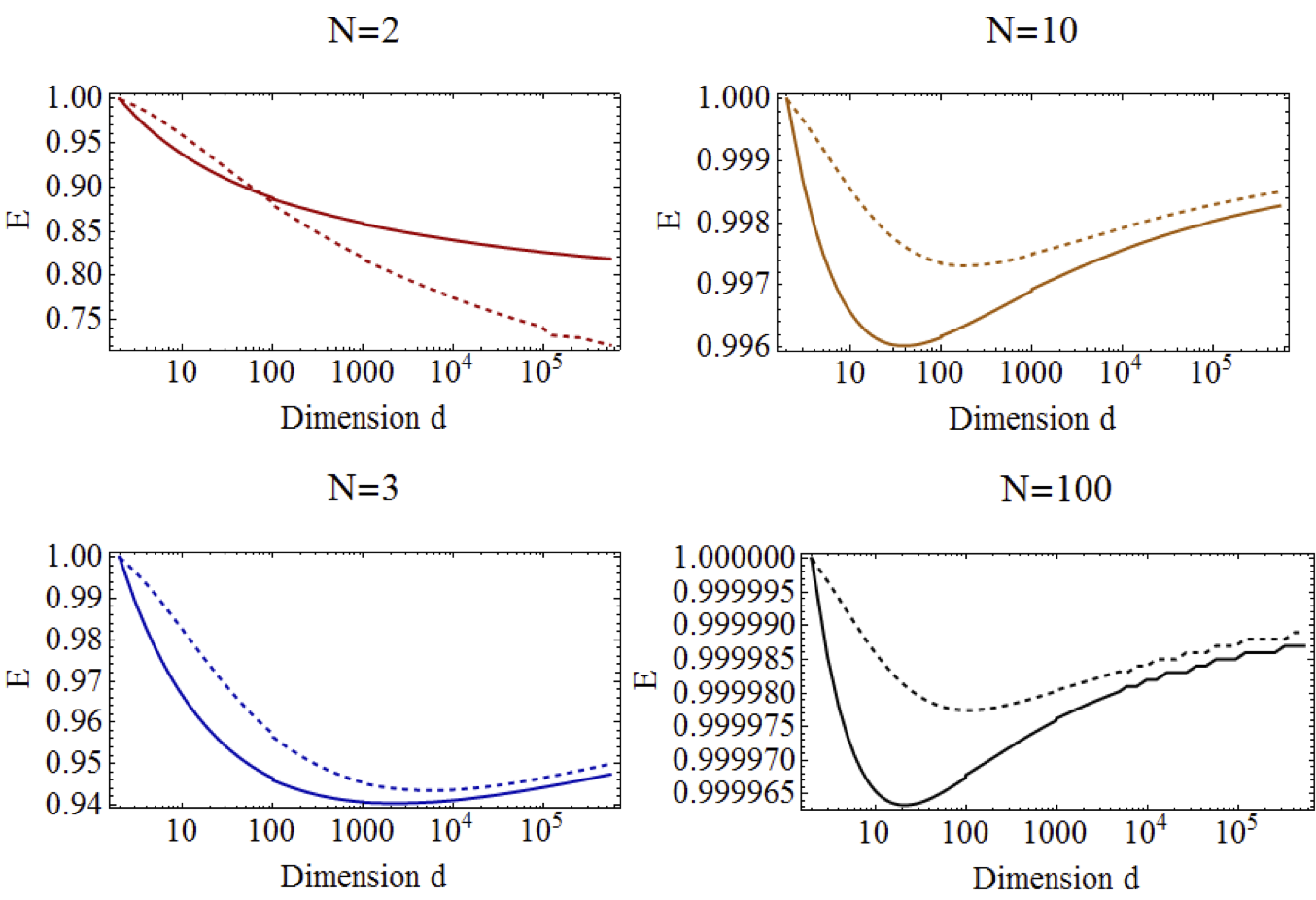}
\caption{The entanglement entropy of the most non-classical state (solid line) and the approximate state $|\Phi_{app}^{(N)}\rangle$ (dashed line) for $N=2,3,10,100$ and dimensions up to $d=0.5\times 10^6$.}
\label{fig:2}
\end{figure}

Using the Schmidt basis representation, the entanglement entropy of a state $|\phi\rangle$ is given by 
\begin{equation} \label{ententropy}
E(|\phi\rangle)=-\sum_{k=0}^{d-1}\lambda_k^2\log(\lambda_k^2)/\log d, 
\end{equation}
which is easily realized to be a number between zero and one (see note \footnote{Note that for two-level systems entropy is usually written in terms of $p\log_2 p=p\log p /\log 2$. Division by $\log 2$ means measuring entropy in units of bits. For $d$-level system, division by $\log d$ corresponds to measuring entropy in units of ``dits''. These are the natural units which normalizes the maximal value of entropy to be 1.} for conventions of units). The entanglement entropy is maximized for the maximally entangled state for which $E(|\psi_{max}\rangle)=1$. 

In Figure \ref{fig:2} we plot the entanglement entropy of the state $|\psi_{opt}\rangle$ and the approximate state $|\Phi_{app}^{(N)}\rangle$ for a few values of $N$ and for dimensions up to about $d=0.5\times 10^6$.

Remarkably, our results show that for the investigated values of $N$ exceeding $N=2$, the entanglement entropy of the most non-classical state is not monotonic with $d$. The sharpness of the trough of the entanglement entropy increases with the value of $N$. Our numerics suggests that $E(|\psi_{opt}\rangle)$ is a non-monotonic function whenever $N\geq3$.

In addition, while for small values of $d$ the state $|\Phi_{app}^{(2)}\rangle$ provides a less good approximation of the entanglement entropy of the most non-classical state \cite{ZRG10}, we see that for $N\geq 3$, $|\Phi_{app}^{(N)}\rangle $ becomes a significantly better approximation of the entanglement entropy of $|\psi_{opt}\rangle$. In fact, the accuracy of the approximate state with respect to $|\psi_{opt}\rangle$ appears to increase with both $N$ and $d$. 
%

Due to the very slow convergence of the entanglement entropy of $|\psi_{opt}\rangle$, especially for fix $N$ and variable $d$, it is hard to tell how it behaves for $d$ beyond the range we have studied. However, it is reasonable to assume that as $d\rightarrow \infty$, the entanglement entropy converges to that of the state $|\Phi_{app}^{(N)}\rangle$. In appendix \ref{AppD}, we show the following limit 
\begin{equation}\label{entropylim}
\lim_{d\rightarrow \infty} E(|\Phi_{app}^{(N)}\rangle)=\left\{
     \begin{array}{lr}
       1/2 & \text{if } N=2 \\
       1 & \text{if } N\geq 3
     \end{array}
   \right.
\end{equation}
Evidently, the entanglement entropy behaves very differently depending on whether we consider case of the CGLMP inequality ($N=2$) or allow for more measurements. Also, this result implies that the entanglement entropy (at the very least for the state $|\Phi_{app}^{(N)}\rangle$) is a non-monotonic function of $d$ whenever $N>2$, which falls in line with our numerical results in Figure \ref{fig:2}.

%
%
The above results show that $(i)$ entanglement entropy of the optimal and approximate state is non-monotonic for $N>2$ and that $(ii)$ it converged to that of the maximally entangled state for $N>2$. Note however, that this does not imply that the optimal state or approximate state are the maximal entangled state for $N>2$ and $d\to\infty$. It simply means that the difference between their corresponding entanglement entropies becomes infinitesimal. An intuitive picture can be obtained by plotting the corresponding Schmidt coefficients $\lambda_i$ as a function of $i$, as was done in \cite{ZG08}. One observes that the shape of the curve is $U$-shaped and that for increasing $d$, as well as $N$ the middle part becomes flatter and flatter. The above result suggests that as $d\to \infty$ the function $\lambda(x)=\lambda_{\lfloor x d \rfloor}$ remains $U$-shaped over $x\in[0,1]$ for $N=2$, while for $N>2$ it becomes essentially a constant function for $x\in(0,1)$, as in the case of the maximally entangled state, apart from a peak with infinitesimal support at $x=0$ and $x=1$. In fact, those two peaks with infinitesimal support make the maximal violations of the Bell inequality for this state differ from that of the maximally entangled state.

Note that entanglement entropy as defined in \eqref{ententropy} is the commonly used measure for entanglement of a state. However, we can also look at the relative entropy or Kullback-Leibler divergence between the maximally entangled state and the optimal or approximate state, i.e.\ $\mathrm{KL}(\psi_{max}| \phi)$ where $|\phi\rangle$ is either the optimal or the approximate state.
\begin{figure}
	\centering
	\includegraphics[width=\columnwidth]{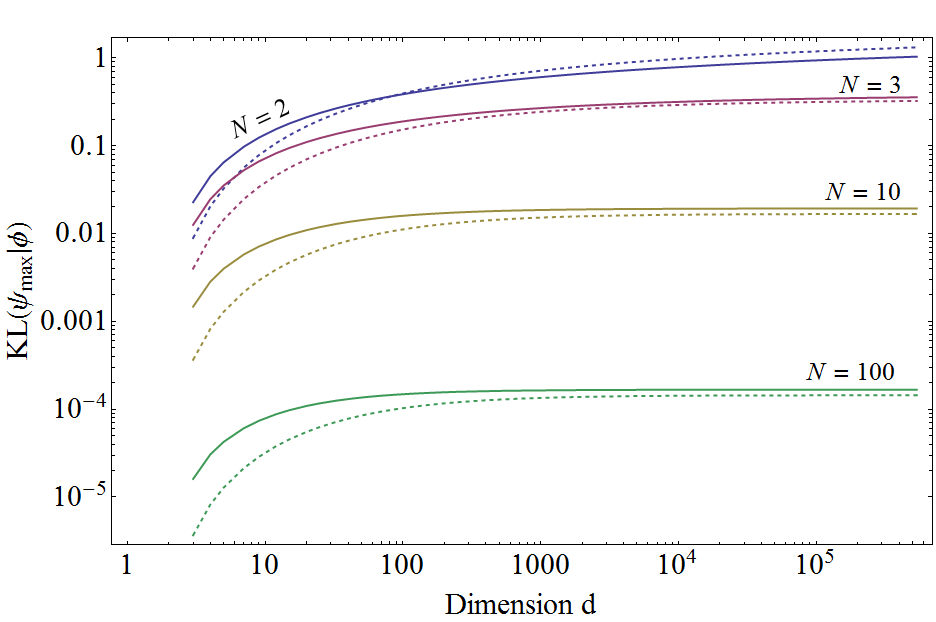}
	\caption{The Kullback-Leibler divergence, for fixed values of $N$, between the maximally entangled state $|\psi_{max}\rangle$ and $|\phi\rangle$, which for solid lines represents the optimal state $|\psi_{opt}\rangle$ and for dashed lines represents the approximate state $|\Phi^{(N)}_{app}$. }
	\label{fig.relent}
\end{figure}
This is shown in Figure \ref{fig.relent} as a function of $d$ for different values of $N$. Furthermore, it is shown in appendix \ref{AppD} that 
\begin{equation} 
\lim_{d\rightarrow \infty} \mathrm{KL}( \psi_{max} | \Phi_{app}^{(N)})=\begin{cases}
  \infty   & \text{$N=2$}, \\
   \log C_N -\frac{4}{N}   & \text{$N\geq 3$},
\end{cases} 
\end{equation}
with $C_N=2^{\frac{4}{N}-1}\sqrt{\pi} \Gamma(1-2/N)/\Gamma(3/2-2/N)$. Note that the divergence for $N=2$ is very slow, $\mathcal{O}(\log\log d)$. This expression agrees with the numerical results shown in Figure \ref{fig.relent}. Furthermore, it is in accordance with the above picture, where we can attribute the divergence of the relative entropy for $N=2$ to a finite difference of the $\lambda_i$ and $1/d$ for infinitely many $i$, corresponding to a finite interval of $x\in[0,1]$, while the finite value of the relative entropy for $N>2$ reflects the fact that $\lambda_i$ and $1/d$ differ only for finitely many $i$, corresponding to the two peaks at $x=0$ and $x=1$ which have infinitesimal support. 

\begin{figure}
	\centering
	\includegraphics[width=\columnwidth]{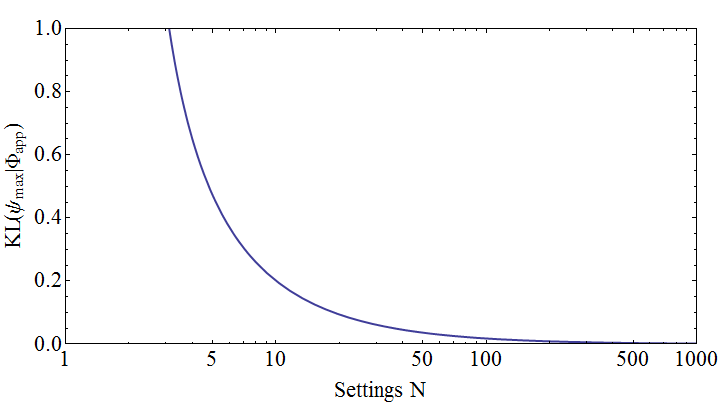}
	\caption{The Kullback-Leibler divergence between the maximally entangled state $|\psi_{max}\rangle$ and the approximate state $|\Phi^{(N)}_{app}\rangle$ as a function of the number of the number of settings $N$ in the limit $d\rightarrow \infty$.}
	\label{fig.infKL}
\end{figure}

To complete the discussion, we plot $\lim_{d\rightarrow \infty} \mathrm{KL}( \psi_{max} | \Phi_{app}^{(N)})$ as a function of $N$ in Figure \ref{fig.infKL}. One sees that the relative entropy between the maximally entangled state and the approximate states for infinite $d$ goes to zero as $N\to\infty$, which is in accordance with the results of \cite{BKP06}, meaning that the peaks of $\lambda(x)$ at $x=0$ and $x=1$ become smaller as $N$ increases until for $N\to\infty$ the approximate state becomes identical to the maximally entangled state.
%
%

\section{Conclusions}
In this work, we have introduced a much simplified version of the mutli-setting generalization of the CGLMP inequality first presented in Ref.\ \cite{BKP06}. We have conjectured the optimal measurements and provided numerics supporting this claim for some low values of $(N,d)$. Then, we studied quantum violations of our inequalities using both maximally entangled and generally entangled states and demonstrated that the maximal violations are obtained by non-maximally entangled states. We gave strong numerical evidence that the Tsirelson bound of the Bell expression $B_{N,d}$ \eqref{Ineq} coincides with the algebraic bound of that quantity, namely zero. However, our main result is the strong evidence for the non-monotonicity of the entanglement entropy of the most non-classical state, as a function of $d$ for $N>2$. 

We see that the most non-classical state for any $N>2$ become more and more entangled with increasing $d$ and its entanglement entropy converges to that of the maximally entangled state. 
This is very much opposite to the quantum behavior observed for the CGLMP inequality \cite{ZRG10} i.e.\ when $N=2$. Our results suggest that the discrepancy between entanglement and the strength of quantum correlations is strongly dependent on the number of settings of the Bell scenario and is eradicated in the limit of large $d$, given more than two settings. Finally, recent work \cite{AZ15} studying the ability of random pure states to violate the CGLMP inequality has shown that with increasing $d$, the inequality is increasingly violated on average. Extending such an analysis to the case of $N>2$ settings would be of interest.




\begin{acknowledgments}
We thank Richard Gill for discussions. Also, we would like to acknowledge the use of the Advanced Research Computing (ARC), University of Oxford in carrying out this work. S.Z. acknowledges support by Nokia Technologies, Lockheed Martin and the University of Oxford through the Quantum Optimisation and Machine Learning (QuOpaL) Project. M.P. was funded by FNP prgramme TEAM and NCN grant 2014/14/E/ST2/00020.
\end{acknowledgments}


\begin{appendix}

\section{Computing the Bell Operator}\label{AppA}
Here we derive Bell operator, $M$, from equation \eqref{M}, associated to the measurements \eqref{Meas1} and \eqref{Meas2}.

Given any shared quantum state written in the Schmidt basis as $|\phi\rangle=\sum_{k=0}^{d-1}\lambda_k|kk\rangle$, the associated probability distribution $P_{N,d}(a,b|x,y)$ can be simplified to
\begin{equation}\label{Prob1}
P_{N,d}(a,b|x,y)\!=\!\frac{1}{d^2}\!\!\sum_{k,l=0}^{d-1}\!\omega^{(k-l)(a-b+\alpha_x^{(N)}+\beta_y^{(N)})}\lambda_k\lambda_{l}.
\end{equation}
We can use the probability distribution to evaluate the Bell expression in equation \eqref{Ineq}. If we additionally observe that $\alpha_n^{(N)}+\beta_n^{(N)}=\frac{1}{2N}$, $\alpha_n^{(N)}+\beta_{n+1}^{(N)}=-\frac{1}{2N}$ and $\alpha_{N-1}^{(N)}+\beta_0^{(N)}=1-\frac{1}{2N}$, one sees that all probabilities appearing in \eqref{Ineq} are independent of the measurement choices. By direct insertion, we write the Bell expression as 
\begin{multline}\label{Prob}
B_{N,d}=\frac{1}{d^2}\sum_{k,l=0}^{d-1}\lambda_{k}\lambda_{l}\Bigg[ \sum_{j=0}^{d-1}(d-j)\omega^{(k-l)\left(j+1-\frac{1}{2N}\right)}+\\ (N-1)\sum_{j=1}^{d-1}(d-j)\omega^{(k-l)\left(j-\frac{1}{2N}\right)}\\
+N \sum_{j=1}^{d-1}(d-j)\omega^{(k-l)\left(-j+\frac{1}{2N}\right)}\Bigg]
\end{multline}
where the Bell operator $M$ appears in the square brackets (scaled by $d^2$).

In case of $k=l$, corresponding to the diagonal elements of $M$, it is easily found that 
\begin{equation}\label{diagonal}
M_{kk}=N-\frac{N-1}{d}.
\end{equation}
To evaluate the off-diagonal elements, corresponding to $k-l\neq 0$, we first use the geometric series to evaluate each of the three series appearing inside the bracket of \eqref{Prob}. Combining the exponentials in trigonometric functions leads to the final expression 
\begin{equation}
B_{N,d}=\sum_{k,l=0}^{d-1}\lambda_k\lambda_l\Bigg[N\delta_{kl}-\frac{N}{d}\frac{\sin\left(\frac{(N-1)\pi (k-l)}{dN}\right)}{\sin\left(\frac{\pi (k-l)}{d}\right)}\Bigg]
\end{equation}
where one identifies the matrix $M$ with the term inside the bracket.

\section{Violations from Maximally entangled states as $d\rightarrow \infty$}\label{AppB}
Here we compute the limit $B_{N,\infty}(|\psi_{max}\rangle)$ by evaluating the integral in equation \eqref{Int}. To compute the integral, we use the Dirichlet series for the cosecant function, namely
\begin{equation}
\frac{1}{\sin\left(z\right)}=-2i\sum_{j=1}^{\infty}e^{iz(2j-1)}.
\end{equation}
If we define $z=\pi(x-y)$, we obtain an expansion for one of the two factors appearing in the integrand of equation \eqref{Int} in terms of the variable $z$. Additionally inferring the notation $f_1=(N-1)\pi (x-y)/N$ and $f_2=(2j-1)\pi (x-y)$, we can write the integral expression in \eqref{Int} as a sum over an infinite series of integrals over integrands of the form $\sin\left(f_1\right)e^{if_2}$. Using Euler's formula, we can write each integral in the infinite series as a sum of four elementary integrals which can easily be solved; each over a term of the form $\cos\left(f_1\pm f_2\right)$ or $\sin\left(f_1\pm f_2\right)$. However, two of the four integrals will vanish: it is a straightforward computation to find that $\int_{0}^{1}\int_{0}^{1}\sin\left(f_1\pm f_2\right)dxdy=0$ for all values of $j$.   



The task of computing $B_{N,\infty}$ is reduced to evaluating the following series of integrals: $N-N\sum_{j=1}^{\infty}\left[\int_{0}^{1}\int_{0}^{1}\left(\cos\left(f_1-f_2\right)
-\cos\left(f_1+f_2\right)\right)dxdy\right]$.
The two elementary integrals appearing in this expression are evaluated to
\begin{equation}
N\int_{0}^{1}\int_{0}^{1}\cos\left(f_1\pm f_2\right)dxdy=\frac{2N^3\left(1-\cos\left(\frac{\pi}{N}\right)\right)}{\pi^2\left(1\mp 2j_\pm N\right)^2}
\end{equation}
where we have defined $j_+=j$ and $j_-=j-1$. Using this expression, one has
\begin{multline}\label{Sum1}
B_{N,\infty}=N+\sum_{j=1}^{\infty} \frac{2N^3\left(1-\cos\left(\frac{\pi}{N}\right)\right)}{\pi^2\left(1-2jN\right)^2}\\
-\sum_{j=1}^{d-1}\frac{2N^3\left(1-\cos\left(\frac{\pi}{N}\right)\right)}{\pi^2\left(1+2(j-1)N\right)^2}.
\end{multline}
The first summation, which we denote $S_1$, can be re-written as  $S_1=N\sin^2\left(\frac{\pi}{2N}\right)/\pi^2\sum_{j=1}^{\infty}\left(j-\frac{1}{2N}\right)^{-2}$.
Recall that the Trigamma function, $\psi_1(z)$, can be defined through the series $\psi_1(z)=\sum_{j=0}^{\infty}(z+j)^{-2}$, which is precisely the form of the summation in $S_1$. If we additionally use the recurrence relation $\psi_1(z+1)=\psi_1(z)-z^{-2}$, $S_1$ can be simplified as 
\begin{equation}
S_1=\frac{N\sin^2\left(\frac{\pi}{2N}\right)}{\pi^2}\psi_1\left(1-\frac{1}{2N}\right).
\end{equation}
A similar analysis can be done for the second sum in \eqref{Sum1}, now denoted $S_2$. First using the recurrence relation, and then using the reflection property of the Trigamma function, $\psi_1(z)+\psi_1(1-z)=\pi^2/\sin^2\left(\pi z\right)$,  allows one to write $S_2=N-N/\pi^2\sin^2\left(\frac{\pi}{2N}\right)\psi_1\left(1-\frac{1}{2N}\right).$

Plugging the final form of $S_1$ and $S_2$ back into \eqref{Sum1}, yields
\begin{equation}
B_{N,\infty}=\frac{2N}{\pi^2}\sin^2\left(\frac{\pi}{2N}\right)\psi_1\left(1-\frac{1}{2N}\right)
\end{equation}
which is the final expression \eqref{Int}.

\section{The numerical technique for finding maximal violations}\label{AppC}

In this paper we have numerically computed both the maximal violation of the inequality \eqref{Ineq} and the entanglement entropy of the associated most non-classical state for values of $d$ ranging up to about $d=0.5\times 10^6$, for a few values of $N$. This was achieved by using similar techniques as in \cite{ZG08} which we will elaborate on here. 

As previously mentioned, the maximal violation of the inequality \eqref{Ineq} is found from computing the smallest eigenvalue of the matrix $M$ \eqref{M}, and the most non-classical state is the associated eigenvector $\vec{\lambda}$ of $M$. Recall that the components of this eigenvector $\vec{\lambda}$ are just the Schmidt coefficients. Since $M$ is a $d\times d$ matrix, the computational complexity of the problem scales as $\mathcal{O}(d^3)$ with increasing $d$, becoming less and less feasible. Nevertheless, there exist more efficient numerical techniques for solving the problem, especially since we are only interested in the smallest eigenvalue of $M$ and the associated eigenvector. 

We have used the method of power iteration over a Krylov subspace which is the essence of the Arnoldi and Lanczos algorithm \cite{Arnoldi, Lanczos}. The Krylov subspace of order $r+1$ associated to some $n\times n$ matrix $A$ and some $n\times 1$ vector $b$ is the space spanned by the image of $b$ under the first $r$ powers of $A$. For large $r$, these images of $b$ approach the eigenvector of $A$ associated to its \textit{largest} eigenvalue (in modulus).

For our purposes, we have fixed $r=20$ and started the iteration from the Schmidt coefficients associated with the maximally entangled state i.e. $\vec{\lambda}^{(0)}=1/\sqrt{d}\times (1,...,1)^T$. However, since we are interested in the \textit{smallest} eigenvalue of $M$, we have decomposed $M$ as
\begin{equation}
M=N I_{d\times d}-M'
\end{equation} 
where $I_{d\times d}$ is the $d$-dimensional identity matrix. We now exploit the simple fact that the eigenvectors of $M$ are the same as the eigenvectors of $M'$ and that the corresponding eigenvalues are related by $m=N-m'$, in particular for $m$ being the smallest eigenvalue of $M$ and $m'$ being the largest eigenvalue of $M'$. 

The above described algorithm is summarized as follows: 

\begin{algorithmic}[1]
\Procedure{Power iteration for optimal state} {}
\State initialize: $\vec{\lambda}^{(0)} \leftarrow  1/\sqrt{d}\times (1,...,1)^T$
\For{$k$ in $1,..,20$}
\State $ \vec{\lambda}^{(k)} = M' \vec{\lambda}^{(k-1)}$
\EndFor
\State normalize: $\vec{\lambda}^{(20)} \leftarrow \vec{\lambda}^{(20)}/||\vec{\lambda}^{(20)}||_2$ 
\State $B=N- (\vec{\lambda}^{(20)})^T M' \vec{\lambda}^{(20)}$ 
\State $E=-\sum_i (\lambda^{(20)}_i)^2 \log_d (\lambda^{(20)}_i)^2$.
\EndProcedure
\end{algorithmic}

Importantly, to avoid memory problems, we only store the current vector $\vec{\lambda}^{(k)}$ in memory when running the iteration. In addition, since $M$ has a banded structure we can store its independent elements in a $d/2$-dimensional vector. Further optimization can be achieved by exploiting the symmetry of $M$ or $M'$. The program is implemented in C++ and ran on the Arcus cluster at the Advanced Research Computing, University of Oxford. Data and code are available on request.

\section{Entanglement entropy of the approximate state $|\Phi_{app}^{(N)}\rangle $}\label{AppD}

In this appendix we derive the limit \eqref{entropylim}. That is, we evaluate  
\begin{equation}
\lim_{d\rightarrow \infty} E(|\Phi_{app}^{(N)}\rangle)= \lim_{d\rightarrow \infty}\left( -\frac{1}{\log d} \sum_{j=0}^{d-1} \lambda_j^2 \log \lambda_j^2 \right)
\end{equation}
where the Schmidt coefficients are given by
\begin{equation}
\lambda_j = \frac{1}{\sqrt{\mathscr{N}}} \frac{1}{\left[(j+1)(d-j)\right]^{\frac{1}{N}}}
\end{equation}
and $\mathscr{N}=\sum_{j=0}^{d-1}\left[(j+1)(d-j)\right]^{-\frac{2}{N}}$ is the normalization. 

We first compute the normalisation to leading order by approximating the sum as an integral. For convenience, we introduce $\epsilon=1/d$ and the rescaled variable $x=j/d$. Then the normalization can be written as
\begin{equation}
\mathscr{N}=\epsilon^{4/N-1} \int_{0}^{1-\epsilon}\frac{dx}{\left[(x+\epsilon)(1-x)\right]^{\frac{2}{N}}} +...,
\end{equation}
where the dots refer to next-to-leading-order terms.
Explicit evaluation of the integral yields to leading order 
\begin{equation} 
\mathscr{N}=\begin{cases}
  2 \epsilon |\log\epsilon| +...   & \text{$N=2$}, \\
   C_N  \epsilon^{4/N-1} + ...   & \text{$N\geq 3$},
\end{cases} \label{eq:Nn}
\end{equation}
where 
\begin{equation}\label{appD:CN}
C_N=2^{\frac{4}{N}-1}\sqrt{\pi}\frac{\Gamma(1-2/N)}{\Gamma(3/2-2/N)}
\end{equation}
which is finite for $N\geq 3$. Note that the explicit form of $C_N$ is not relevant for this derivation, but will be relevant later.

Approximating the sum in the entanglement entropy in a similar manner one arrives at
\begin{multline}\label{ententropy}
E(|\Phi_{app}^{(N)})=\frac{\log\left(\mathscr{N} \epsilon^{-4/N}\right) }{| \log \epsilon |} + \\
+\frac{2}{N}\frac{\epsilon^{4/N-1}}{\mathscr{N}| \log \epsilon |} \int_{0}^{1-\epsilon}\frac{\log\left[(x+\epsilon)(1-x)\right]}{\left[(x+\epsilon)(1-x)\right]^{\frac{2}{N}}}dx + ....
\end{multline}

To take the limit $d\to \infty$, or equivalently $\epsilon\to 0$, of \eqref{ententropy}, we calculate the leading order contribution from both terms. Using \eqref{eq:Nn} one obtains for the first term
\begin{equation}
\frac{\log\left(\mathscr{N} \epsilon^{-4/N}\right) }{| \log \epsilon |} =1+ ... \label{eq:first-term}
\end{equation}

The integral in the second term can be evaluated as follows for $N=2$
\begin{equation}
\int_{0}^{1-\epsilon}\frac{\log\left[(x+\epsilon)(1-x)\right]}{\left[(x+\epsilon)(1-x)\right]^{\frac{2}{N}}}dx = - |\log\epsilon|^2 +...  
\label{eq:int1}
\end{equation}
while for $N\geq 3$ one has 
\begin{equation}
\int_{0}^{1-\epsilon}\frac{\log\left[(x+\epsilon)(1-x)\right]}{\left[(x+\epsilon)(1-x)\right]^{\frac{2}{N}}}dx = 
   \mathcal{O}(\epsilon^{1-2/N} |\log \epsilon| ) \label{eq:int2}
\end{equation}
Combining this with \eqref{eq:Nn}, we see that the second term in \eqref{ententropy} becomes 
\begin{multline}
\frac{2}{N}\frac{\epsilon^{4/N-1}}{\mathscr{N}| \log \epsilon |} \int_{0}^{1-\epsilon}\frac{\log\left[(x+\epsilon)(1-x)\right]}{\left[(x+\epsilon)(1-x)\right]^{\frac{2}{N}}}dx \\
= \begin{cases}
  -\frac{1}{2}+...   & \text{$N=2$}, \\
   \mathcal{O}(\epsilon^{1-2/N}) & \text{$N\geq 3$},
\end{cases} \label{eq:second-term}
\end{multline}
which for $N\geq 3$ becomes zero in the limit $\epsilon\to 0$. Plugging \eqref{eq:first-term} and \eqref{eq:second-term} into \eqref{ententropy} and taking the limit finally gives
 \begin{equation}
\lim_{\epsilon\to 0}E(|\Phi_{app }^{(N)}\rangle)=
 \begin{cases}
  \frac{1}{2} & \text{$N=2$}, \\
  1 & \text{$N\geq 3$},
\end{cases}
\end{equation}
which completes the derivation.

For completeness we also calculate the relative entropy or Kullback-Leibler divergence between the maximally entangled state and the approximate state in the limit where the number of outcomes goes to infinity. That is
\begin{equation}
\lim_{d\rightarrow \infty} \mathrm{KL}( \psi_{max} | \Phi_{app}^{(N)})= \lim_{d\rightarrow \infty}\left( -\frac{1}{d} \sum_{j=0}^{d-1}  \log d \lambda_j^2 \right)
\end{equation}
Following a similar strategy as above one obtains
\begin{eqnarray}
&&\mathrm{KL}( \psi_{max} | \Phi_{app}^{(N)}) =  \nonumber\\ 
&=& \log\left( \mathcal{N}\epsilon^{1-\frac{4}{N}} \right) + \frac{2}{N} \int_0^{1-\epsilon} \log\left[(x+\epsilon)(1-x)\right] dx +...  \nonumber\\
&=&  \log\left( \mathcal{N}\epsilon^{1-\frac{4}{N}} \right) +  \frac{2}{N} (-2 +2\epsilon -2\epsilon\log\epsilon) +...
\end{eqnarray}
In the limit where $d\to\infty$, or equivalently $\epsilon\to0$, the second term converges to $-4/N$. Furthermore, we see from \eqref{eq:Nn} that the first term diverges for $N=2$, while it converges to $\log C_N$ for $N\geq3$. In total one has
\begin{equation} 
\lim_{d\rightarrow \infty} \mathrm{KL}( \psi_{max} | \Phi_{app}^{(N)})=\begin{cases}
  \infty   & \text{$N=2$}, \\
   \log C_N -\frac{4}{N}   & \text{$N\geq 3$},
\end{cases} 
\end{equation}
where $C_N$ is given in \eqref{appD:CN}. Note that the divergence for $N=2$ is very slow. Indeed from the above expression, one sees that it diverges as $\mathcal{O}(\log\log d)$ when $d\to \infty$.

\end{appendix}


\begin{thebibliography}{20}
\bibitem{B64}
J.~S.~Bell, 
On the Einstein-Podolsky-Rosen paradox,
 Physics,	\textbf{1}, 195 (1964).	
	



\bibitem{HHHH09}
R.~Horodecki, P.~Horodecki, M.~Horodecki, and K.~Horodecki,
Quantum entanglement,
Rev. Mod. Phys. \textbf{81}, 865 (2009).

\bibitem{BBP96}
C.~H.~Bennett,  H.~J.~Bernstein, S.~Popescu, and
B.~Schumacher,
Concentrating partial entanglement by local operations,
Phys. Rev. A \textbf{53}, 2046 (1996).



\bibitem{CHSH69}
J.~F.~Clauser, M.~A.~Horne, A.~Shimony, and R.~A.~Holt,
Proposed Experiment to Test Local Hidden-Variable Theories,
Phys. Rev. Lett. \textbf{23}, 880 (1969).

\bibitem{F82}
A.~Fine,
Phys. Rev. Lett. \textbf{48}, 291 (1982).

\bibitem{CGLMP02}
D.~Collins, N.~Gisin, N.~Linden, S.~Massar, S.~Popescu,
Bell Inequalities for Arbitrarily High-Dimensional Systems,
Phys. Rev. Lett. \textbf{88}, 040404 (2002).



\bibitem{AGG05}
A.~Acin, R.~Gill, and N.~Gisin,
Optimal Bell Tests Do Not Require Maximally Entangled States,
Phys. Rev. Lett. \textbf{95}, 210402 (2005). 


\bibitem{ADGL02}
A.~Acin, T.~Durt, N.~Gisin and J.~I.~Latorre,
Quantum nonlocality in two three-level systems,
Phys. Rev. A \textbf{65}, 052325 (2002).

\bibitem{ZG08}
S.~Zohren and R.~D.~Gill,
Maximal Violation of the Collins-Gisin-Linden-Massar-Popescu Inequality for Infinite Dimensional States,
Phys. Rev. Lett. \textbf{100}, 120406 (2008).


\bibitem{FP15}
E.~A.~Fonseca and F.~Parisio,
There is no anomaly in the nonlocality of two entangled qutrits,
arXiv: 1504.01601.


\bibitem{joinedreference}
S.~Braunstein and C.~Caves,
Writing out better Bell inequalities,
Ann. Phys. (NY), \textbf{202}, 22 (1990).;
N.~Gisin,
Bell inequality for arbitrary many settings of the analyzers,
Phys. Lett. A, \textbf{260}, 1 (1999).;
I.~Pitowsky and K.~Svozil,
Optimal tests of quantum nonlocality,
Phys. Rev. A \textbf{64}, 014102 (2001).;
N.~Brunner and N.~Gisin,
Partial list of bipartite Bell inequalities with four binary settings,
Phys. Lett. A, \textbf{372}, 3162 (2008).;
K.~F.~Pal and T.~Vertesi,
Quantum bounds on Bell inequalities,
Phys. Rev. A \textbf{79}, 022120 (2009);
K.~F.~Pal and T.~Vertesi,
Maximal violation of a bipartite three-setting, two-outcome Bell inequality using infinite-dimensional quantum systems, 
Phys. Rev. A \textbf{82}, 022116 (2010).

\bibitem{CG04}
D.~Collins and N.~Gisin,
A relevant two qubit Bell inequality inequivalent to the CHSH inequality,
J. Phys. A. Math. Gen. \textbf{37}, 1775 (2004).


\bibitem{joinedreference2}
S-W.~Ji, J.~Lee, J.~Lim, K.~Nagata, and H-W.~Lee,
 Multisetting Bell inequality for qudits,
Phys. Rev. A \textbf{78}, 052103 (2008).;
M. Zukowski and A. Datta,
Geometric chained inequalities for higher-dimensional systems,
Phys. Rev. A \textbf{90}, 012106 (2014).

\bibitem{L08}
E.~R.~Loubenets,
Multipartite Bell-type inequalities for arbitrary numbers of settings and outcomes per site,
J. Phys. A: Math. Theor \textbf{41}, 445304 (2008).




\bibitem{BKP06}
J. Barrett, A. Kent and S. Pironio,
Maximally Non-Local and Monogamous Quantum Correlations,
Phys. Rev. Lett. \textbf{97}, 170409 (2006).


\bibitem{VB96}
L. Vandenberghe and S. Boyd, SIAM Review \textbf{38}, 49 (1996).


\bibitem{CC06}
J-L. Chen, C. Wu, L. C. Kwek, C. H. Oh, and M-L. Ge,
Violating Bell inequalities maximally for two d-dimensional systems,
Phys. Rev. A \textbf{74}, 032106 (2006).


\bibitem{ZRG10}
S. Zohren, P. Reska, R. D. Gill and W. Westra,
A tight Tsirelson inequality for infinitely many outcomes,
EPL \textbf{90}, 10002 (2010).



\bibitem{AZ15}
M. R. Atkin and S. Zohren,
Violations of Bell inequalities from random pure states,
arXiv 1407.8233


\bibitem{Arnoldi}
W. E. Arnoldi, 
Quarterly of Applied Mathematics, volume 9, pages 17–29, 1951.

\bibitem{Lanczos}
C. Lanczos, 
J. Res. Nat’l Bur. Std. \textbf{45}, 225-282 (1950).



\end{thebibliography}
\end{document}